\documentclass{IOS-Book-Article}     
\usepackage{xurl}

\begin{document}
\begin{frontmatter}          
%
\title{Dialogues with algorithms}

\author{
\snm{Joost J. Joosten}
\orcid{0000-0001-9590-5045}
\thanks{Postal address: Joost J. Joosten, Faculty of Philosophy, Carrer Montalegre 6, 08001, Barcelona, Catalonia, Spain. e-mail: \url{jjoosten@ub.edu}. Website: \url{http://www.joostjjoosten.nl//}}}
\address{University of Barcelona}
\runningauthor{}
%
%
\begin{abstract}
In this short paper we focus on human in the loop for rule-based software used for law enforcement. For example, one can think of software that computes fines like tachograph software, software that prepares evidence like DNA sequencing software or social profiling software to patrol in high-risk zones, among others. 

An important difference between a legal human agent and a software application lies in possible dialogues. A human agent can be interrogated to motivate her decisions. Often such dialogues with software are at the best extremely hard but mostly impossible.

We observe that the absence of a dialogue can sincerely violate civil rights and legal principles like, for example, Transparency or Contestability. Thus, possible dialogues with legal algorithms are at the least highly desirable. Futuristic as this may sound, we observe that in various realms of formal methods, such dialogues are easily obtainable. However, this triggers the usual tension between the expressibility of the dialogue language and the feasibility of the corresponding computations.
\end{abstract}


\end{frontmatter}


\section{Human in the loop}
It is commonly held that during automated legal decision making there should be human oversight and involvement. As a matter of fact, the human involvement is anchored in various legal instruments most notably in the European GDPR \cite{GDPR}, quoting from Article 22.1:
\begin{quote}
The data subject shall have the right not to be subject to a decision based solely on automated processing, including profiling, which produces legal effects concerning him or her or similarly significantly affects him or her.
\end{quote}
The more recent European AI Act dedicates an entire article to human oversight,
quoting from Article 14.1 (Human oversight):
\begin{quote}
High-risk AI systems shall be designed and developed in such a way, including with
appropriate human-machine interface tools, that they can be effectively overseen by
natural persons during the period in which the AI system is in use.
\end{quote}

However, it is far from clear what such human involvement should look like. Some argue that often, having a human in the loop is considered unjustly as a magic potion to warrant correct decisions but that human oversight in general falls short as a solution to the risks of algorithmic decision-making (\cite{FalseComfort}). Notwithstanding, all scholars agree that some degree of human oversight and involvement is needed though, again,  there is no common notion of what this should look like (see also \cite{HumanOversight}). Human overview should minimise or mitigate undesired effects of using AI and automated decision making like biases/discrimination, nudging, opaqueness or simply errors (\cite{WeertsEtAl, NISTDoc, Yeung}).

An important parameter in the discussion on correct human involvement in legal automated decision making is the kind of AI that is used. For example, transparency in Neural Networks is much harder to achieve (if not impossible) than in old-style rule based AI. In this paper we shall therefore focus on the latter since, as we shall see, recent development of formal methods can facilitate certain rudimentary forms of interrogations of algorithms on how they perform and what kind of properties these algorithms have. 

\section{Black-boxes and dialogues}

We decide to focus on classical legal computer programs in this paper leaving other paradigms like neural networks and the like aside. Thus, we focus on programs that follow our human ideal logical reasoning schemes in an algorithmic fashion. The resulting legal computer programs often leads to problems. Users or those affected may object to legal software and claim to lose transparency, oversight, understanding and fear errors. As a matter of fact, all substantially large computer programs do contain errors, be they typos, small design errors, or programmed biases. However, all these objections also holds for human actors in the law enforcement who also err, have personal inclinations and preferences and may be methodologically far from optimal. 

An essential difference between legal computer programs and human legal actors lies in the possibility to \emph{dialogue}. For example, with a human legal actor there is less fear of loss of transparency. For sure, the legal actor may and most likely will be much more knowledgeable than us, but at least we have the feeling that we can pose questions, ask explanations, and inquire for the basic assumptions.  Those acts of interaction are at least cumbersome with a program and not accessible to the average citizen. In various known cases of erroneous software in the past we see that interaction took very long, and over sometimes years went through various committees and groups of experts. 

A notorious example is the Dutch SyRI social risk scoring computer algorithm (\cite{SyRI}). SyRI would assign a fraud risk score to citizens on the basis of which social support could be denied. Only after years of allegations and human tragedy of involved individuals it was shown and understood that the algorithm was biased and, for example, would not act equally on citizens that have more than one nationality. We imagine the amount of tragedy and time that could have been saved if at an early stage, one could have asked the program how it functioned. For example, imagine that $x$ and $y$ range over data containers for individual citizens. And imagine that $\mathcal A$ is the set of all attributes\footnote{For the sake of exposition we restrict to unary attributes, like ``$z$ is female"  but the example can easily be extended to relations of higher arity.} of those citizens, that ${\sf NrOfPassports}(z)$ is a function that tells how many different passports (nationalities) an individual $z$ possesses and that ${\sf Score} (z)$ is the score that individual $z$ obtains by the SyRI algorithm. We can then formalise the question of whether SyRi would be biased for multiple passport holders:
\[
\begin{array}{ll}
\forall x, y \ \Big( ({\sf NrOfPassports}(x) \neq {\sf NrOfPassports}(y)) \wedge & \bigwedge_{P\in \mathcal A} (P(x) \leftrightarrow P(y)) \\
 & \ \ \Longrightarrow \ \ {\sf Score}(x) = {\sf Score}(y)\Big)\\
\end{array}
\]
Clearly, this is a property that either holds true or false of the SyRI software. If only citizens would have been able at an early stage to query this question to the program, it is likely that the whole affair would have been less painful and time-consuming. In general, one can imagine that enabling dialogues with programs could restore trust and control in the interaction between human actors and software. 

We wish to stress that merely having access to the source code of a legal program is not a sufficient condition to gain transparency. Even for IT specialists it is extremely hard to fully understand the exact working of computer code. As a matter of fact in its full generality, full understanding of the source code is impossible since it would imply that we can solve the unsolvable Halting Problem.

However, it seems like a bare minimum to at least grant access to the source code so that citizens that are affected by the functioning of that code may try to understand how the code works. In this regard it is curious to mention the BOSCO case in Spain. BOSCO is a state owned computer program that decides who qualifies for social financial support. Supposedly, BOSCO follows a fully determined legal text but numerous wrong judgements made by the program have been reported \cite{BOSCO}. Notwithstanding supported claims of errors, Spanish administration is still reluctant to disclose the software. This is to be contrasted with French practice and regulation \cite{FRANCE}, where they strive for open software in public administrations.

Ponce-Solé points out in \cite{Ponce} that Article 9.3 of the Spanish Constitution prohibits arbitrariness in legal decision making. The very same article however also mentions that norms should be publicly announced. This begs the discussion that if the implementation of the law fills in substantial blanks in that law or reinterprets a law, if it then should be allowed for this implementation to be proprietary or to remain undisclosed. Note that filling blanks or reinterpreting is typically needed to go from natural language to executable code.

Through the examples above, we think it is convincingly showcased that when there is no access to nor precise understanding of the source code, transparency is at stake and it will become extremely hard for citizens to contest automated decisions that concern their rights.

\section{Formal methods enable dialogues}
Engaging in a dialogue with a software program may sound extremely futuristic. In a rudimentary form, however, this is possible and almost already in place. However, it can only be applied in the case the software is embedded in an environment of formal methods like software correctness proofs or model checking. Formal methods refers to a large collection of techniques where mathematics and logic is employed to reason about, most prominently, the correctness of algorithms/software. As we have seen, correctness/robustness is of utmost importance and currently receives much attention. Robustness is discussed in the European AI Act (e.g.~\cite{AIAct}, article 15) and it seems that the only way to achieve a serious level of robustness is by employing formal methods. Often, the use of formal methods implicitly opens the door to enabling rudimentary dialogues with software. We shall discuss this for two paradigms: model checking and software synthesis through proof assistants. Let us start with the latter.

Proof assistants typically check mathematical proofs for correctness. One may think, aren't mathematical proofs by definition correct? The answer is no. That is to say, most proofs will contain minor errors though oftentimes these errors can easily be repaired by slightly tweaking the argument. Sometimes, mathematicians don't even see the small error since it is clear how the global logical structure of the argument goes. Another `error' could be the omission of an easy reasoning step. Proof assistants like Coq, Isabelle or Lean, to mention only a few, are small computer programs that perform a simple task. When they are presented with a mathematical proof, they will check step by step that each alleged larger-scale reasoning indeed has a proof. Using proof assistants in mathematics has lead to various new insights, a few new theorems and numerous detections of flaws in proofs. \cite{Geuvers}

If the language of a proof assistant is rich enough, one can express software in it and moreover, one can express substantial software \emph{behaviour} in it. Thus, in the environment of a proof assistant, one can make claims about the software. We call this a formal specification when the claims fully describe the desired behaviour of the software. Consequently, once a piece of software lives inside a proof assistant environment, this automatically enables questions to be posed about the software and behold our dialogue. The caveat here is that it will be the programmer/user of the proof assistant who will need to provide a formally verified answer to the question. This means answering the question and proving that the answer is indeed correct. This feels like falling short as a real dialogue but at least certainty about the answers will be obtained (provided we accept that the very small proof assistant program itself is correct). Moreover, very few software is being obtained through the use of proof assistants, let alone legal software. In this context we mention a project to formalise European (freight) traffic regulation software inside Coq that has resulted so far to a formally verified time library \cite{FVTIME}.

Within model checking in law, it seems that dialogues may come somewhat easier. A legal model checking paradigm is described in \cite{ModelCheck561} and would run as follows. In this paradigm, a computable law would be expressed as a formula $\varphi$ in some formal language $\mathcal L$ that is rich enough to express the law under consideration. Next, we consider data files that describe particular cases to which the law should be tested. The data files are formally viewed as mathematical structures often called model. Thus, we can consider each data file as a model $\mathcal M$ and a different data file gives rise to a (typically) different model. If we wish to inquire if the case $\mathcal M$ is legal or not according to the law $\varphi$ we resort to techniques of model checking and the question will boil down to 
\[
\mathcal M \models \varphi \ \ ?
\]
Thus, given a model $\mathcal M$ and a formula $\varphi$, does the model $\mathcal M$ make true the formula $\varphi$ yes or no?
We should stress here that this question in general need not be decidable (recall the Halting Problem). Or if it is decidable, it may not be feasible. The art in legal model checking thus resides in choosing the language $\mathcal L$ rich enough so that various interesting laws can be expressed in it. On the other hand, the language should not be too rich as to prevent undecidability or unfeasibility to kick in. Once such a balance is found, there are certain benefits of model checking over proof assistants: the same model-checking framework will work for a whole class of laws, whereas just minor tweaks in legal formally verified software may imply enormous tasks for the programmer to generate new proofs. 

Of course, a model checker does not directly yield error free software since the implementation may still contain errors. An optimal situation seems to arise if the model checking algorithm is implemented using proof assistants but let us leave this matter aside here.

One can also consider the \emph{consistency question}: is the law $\varphi$ consistent, that is, is there some situation/model $\mathcal M$ that abides by the law, that is, is there some $\mathcal M$ so that $\mathcal M\models \varphi$? Directly related to the consistency question is the \emph{tautology question}: is the law $\varphi$ true in all possible situations/models?  We use the standard notation 
\[
\models \varphi
\]
for the tautology statement: $\varphi$ holds true on all models $\mathcal M$. In our setting this is tantamount to saying that the law $\varphi$ is satisfied in every possible situation $\mathcal M$. It must be observed that the question $\models \varphi$ looks more complicated than $\mathcal M \models \varphi$ for a particular model. In general, this holds true and the tautology question is really harder (where the notion of harder being strictly harder often depends on complexity questions like ${\sf P} = {\sf NP}$) than just the model checking question. Notwithstanding, for various logics, like Linear Temporal Logic, the corresponding tautology question is decidable with not too bad computational properties.

Using the tautology question we can now enter in dialogue with the law as long as the dialogue is restricted to the linguistic fragment $\mathcal L$. Let $\psi$ be some property that can be expressed in $\mathcal L$. The question of whether applying the law $\varphi$ necessarily leads to having the property $\psi$ can thus be stated as 
\[
\models \varphi \to \psi .
\]
We observe the difference in both paradigms: in the proof assistant environment we could directly ask questions about the software. However, these questions were to be replied and proven by the user itself. In the model checking environment, we can pose questions about the law $\varphi$. In this case, however, the questions are automatically answered by the model (tautology) checking algorithm. One can argue that a formalisation $\varphi$ of a computable law can actually be seen as a program. Up to now, laws are typically written in natural languages and a formalisation $\varphi$ in a logic $\mathcal L$ can be seen as a program: a translation of a written computable law into a particular model of computation with the formal specification being quite similar to a program. 



\end{document}